\documentclass[fleqn,usenatbib]{mnras}

\usepackage{newtxtext,newtxmath}

\usepackage[T1]{fontenc}
\DeclareRobustCommand{\VAN}[3]{#2}
\let\VANthebibliography\thebibliography
\def\thebibliography{\DeclareRobustCommand{\VAN}[3]{##3}\VANthebibliography}

\usepackage{graphicx}	
\usepackage{amsmath}	

\DeclareMathOperator{\sech}{sech}

\title[How can LISA probe GW190425-like binary neutron stars?]{How can LISA probe a population of GW190425-like binary neutron stars in the Milky Way?}

\author[V. Korol et al.]{
Valeriya Korol,$^{1}$\thanks{E-mail: korol@star.sr.bham.ac.uk}
and Mohammadtaher Safarzadeh$^{2}$
\\
$^{1}$School of Physics and Astronomy \& Institute for Gravitational Wave Astronomy, University of Birmingham, Edgbaston, Birmingham B15 2TT, UK\\
$^{2}$Department of Astronomy and Astrophysics, University of California, Santa Cruz, CA 95064, USA
}

\date{Accepted 1 February. Received 5 December; in original form ZZZ}

\pubyear{2020}

\begin{document}
\label{firstpage}
\pagerange{\pageref{firstpage}--\pageref{lastpage}}
\maketitle

\begin{abstract}
The nature of GW190425, a presumed binary neutron star (BNS) merger detected by the LIGO/Virgo Scientific Collaboration (LVC) with a total mass of $3.4^{+0.3}_{-0.1}$ M$_{\odot}$, remains a mystery.
With such a large total mass, GW190425 stands at five standard deviations away from the total mass distribution of Galactic BNSs of $2.66\pm 0.12$ M$_{\odot}$. LVC suggested that this system could be a BNS formed from a fast-merging channel rendering its non-detection at radio wavelengths due to selection effects. BNSs with orbital periods less than a few hours -- progenitors of LIGO/Virgo mergers -- are prime target candidates for the future Laser Interferometer Space Antenna (LISA).
If GW190425-like binaries exist in the Milky Way, LISA will detect them within the volume of our Galaxy and will measure their chirp masses to better than 10\,per cent for those binaries with gravitational wave frequencies larger than 2\,mHz. 
This work explores how we can probe a population of Galactic GW190425-like BNSs with LISA and investigate their origin.
We assume that the Milky Way's BNS population consists of two distinct sub-populations: a fraction $w_1$ that follows the observed Galactic BNS chirp mass distribution and $w_2$ that resembles chirp mass of GW190425. We show that LISA's accuracy on recovering the fraction of GW190425-like binaries depends on the BNS merger rate. For the merger rates reported in the literature, $21 - 212\,$Myr$^{-1}$, the error on the recovered fractions varies between $\sim 30 - 5$\,per cent.
\end{abstract}




\section{Introduction}
\label{sec:intro}

GW190425 is a compact object merger with a total mass of $3.4^{+0.3}_{-0.1}$\,M$_{\odot}$, that was recently detected by the LIGO/Virgo Collaboration \citep[LVC, ][]{gw190425}. 
If GW190425 is a binary neutron star (BNS) merger, its total mass is inconsistent with the observed Galactic BNS population that shows a narrow range in the total mass of $\approx 2.66\pm 0.12$\,M$_{\odot}$ \citep{far19}.
LVC suggests that GW190425 might belong to a class of BNSs born from a fast-merging channel. 
In this hypothesis, massive BNSs like GW190425 merge on timescales of 10 - 100\,Myr due to being formed with short orbital periods through unstable
case-BB mass transfer or with high eccentricities through large natal
kicks \citep[e.g.][]{tau17,2020MNRAS.496L..64R,gal20}. 
Short life-times and severe Doppler smearing, which affects short-period systems, make these binaries invisible to radio telescopes and thus hard to find in our Galaxy \citep{cam18,Pol2020}.

However, the comparable merger rates of GW190425 with $\mathcal{R}_{\rm GW190425}=460^{+1050}_{-390}$\,yr$^{-1}$\,Gpc$^{-3}$ and that of GW170817 with $\mathcal{R}_{\rm GW170817}=760^{+1740}_{-650}\,{\rm yr^{-1}\,Gpc^{-3}}$ derived by LVC can be challenging to account for through a fast-merging channel hypothesis for two reasons.  First, massive neutron stars are expected to form from more massive progenitor stars, thus if adopting the initial mass function of \citet{sal55}, the expected merger rate of GW190425-like systems should result in a lower merger rate for GW190425 compared to GW170817 \citep[for a more detailed discussion see][]{saf20}. Second, if fast merging BNSs form through unstable case-BB mass transfer \citep[e.g.][]{iva03,dew03}, they should constitute to $\lesssim 10\,$per cent of the total BNS number according to a suite of simulations studied in \citet{saf19}. In addition, a large fraction of the BNSs formed through fast-merging channels could challenge the BNS origin for the r-process enrichment of the ultra-faint dwarf galaxies \citep[e.g.][]{kom14,mat14,saf17,saf19}.

Many binary population synthesis studies investigated the observed properties of Galactic BNSs \citep[for an overview see][]{tau17}.
Although these studies are able to reproduce the broad characteristics of the BNS population, they seem to have difficulties in matching the mass distribution of the radio population and simultaneously account for the unusually high mass of GW190425 \citep[e.g.][see also Fig.~\ref{fig:Mchirp_distribution}]{tau17,vig18,kru20}.
Binary population synthesis models using probabilistic remnant mass and kicks prescriptions form heavier BNSs, still half of their population have masses larger than those of known Galactic BNSs \citep{2021MNRAS.500.1380M}. Thus, it is not yet clear if more GW190425-like BNS could be associated with the fast-merging channel or whether they require different explanations.
For example, massive BNSs may not be fast-merging at all, instead the lack of radio detections could be attributed to a weak neutron star magnetic dipole moment. If pulsars are born either with a very strong or extremely weak magnetic dipole moment, they will migrate into the graveyard of pulsars and become invisible to radio telescopes \citep{saf20}.

Regardless of the origin, if massive short-period GW190425-like binaries exist in the Milky Way, the Laser Interferometer Space Antenna \citep[LISA,][]{lisa} will be able to detect them and - for sufficiently short orbital periods - accurately measure their chirp masses. 
In this work we test whether LISA can identify massive BNSs as a distinct Galactic sub-population.
We assume that two sub-populations of BNSs reside in the Milky Way: one that follows the chirp mass distribution of known Galactic BNSs, and another one that resembles GW190425. The question at hand is that given the expected merger rate of BNSs in the Milky Way, can LISA detect short period massive BNSs and recover their true fraction?

The idea that LISA can study the existence of fast-merging channel BNS system has been introduced in other works. Recently, \citet{and20} showed that even by adopting a pessimistic Milky Way merger rate for Galactic BNS systems, LISA would detect tens of such BNSs within four years of observations. \citet{lau20} arrived at similar conclusions suggesting that with the achievable high accuracy on the chirp mass and eccentricity, one can constrain BNS formation scenarios such as the natal kicks imparted to neutron stars at birth. In this work we quantify how many events it would take for LISA to set bounds on the fraction of GW190425-like systems in the Milky Way. 
It is important to mention that in this work, we do not make any assumptions regarding BNS GW frequencies. 
If GW190425-like systems come from the fast-merging channel, it is possible that they would appear at higher frequencies in the LISA band compared to the bulk of the Milky Way population. 
Alternatively, they could form at similar frequencies as the currently known Galactic BNSs but with high eccentricities. 
Thus, we anticipate that to associate massive BNSs with the fast-merging channel additional measurements of the frequency and eccentricity distributions would be required.

The structure of this work is as follows. In Section 2, we analyse LISA's capability to determine the fractional error on the chirp mass of binaries as a function of the gravitational wave (GW) frequency. 
In Section 3, we set up a multivariate Gaussian distribution for the binaries detectable by LISA, one following the chirp mass distribution of the Galactic binaries and the other resembling GW190425. We study how LISA can recover the relative fraction of the two sub-populations. In Section 4, we discuss our results and conclude. 

\section{Methods}

\subsection{Chirp mass measurement}

\begin{figure*}
\centering
\includegraphics[width=1.5\columnwidth]{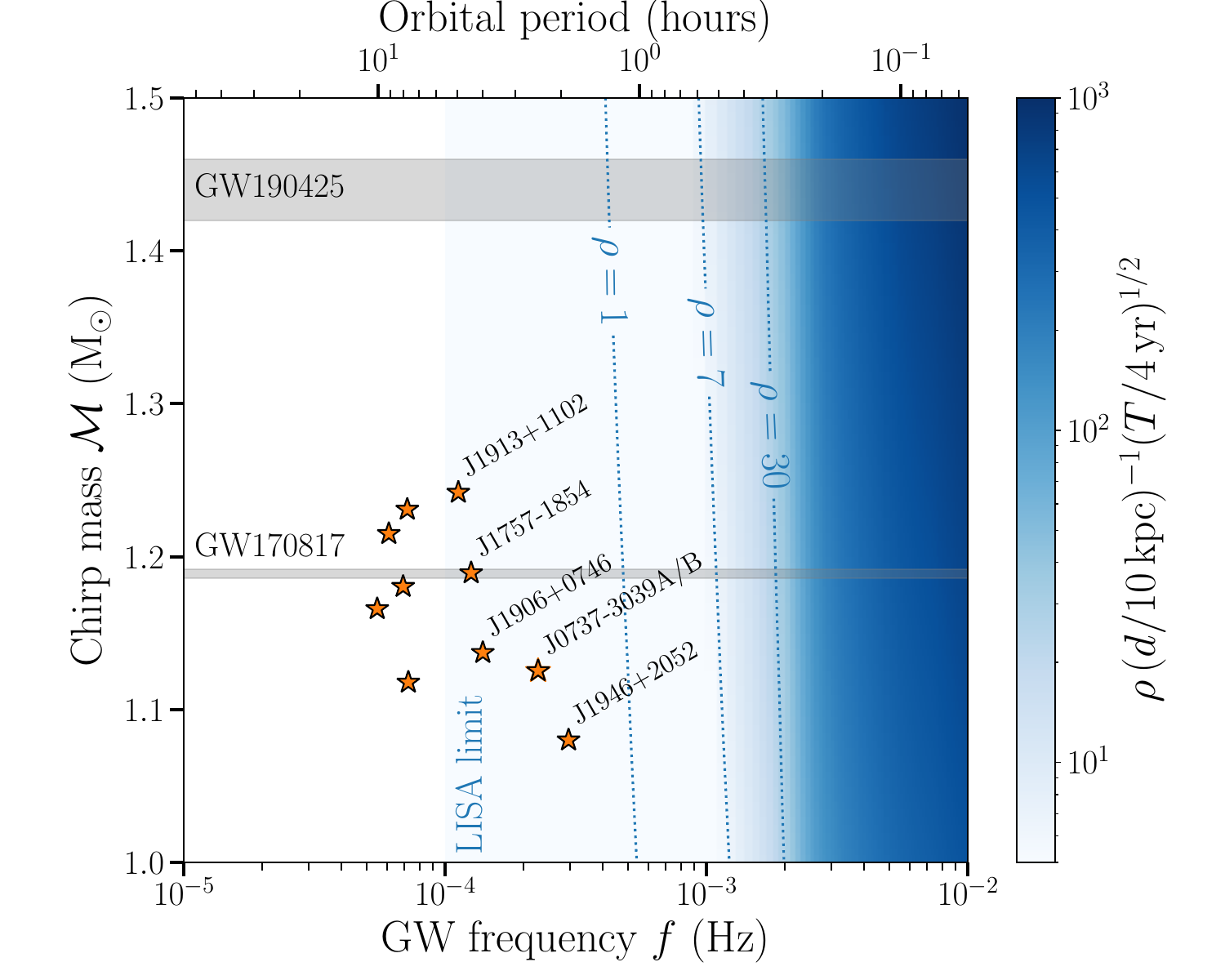}
\caption{Detectability of Galactic BNSs in the chirp mass-frequency parameter space. In colour, we show the sky-, inclination- and polarisation-averaged SNR for a circular BNS placed at the distance of $10\,$kpc after nominal four years of LISA mission. Dotted contours show SNR=1 representing the instrument noise, SNR=7 representing the nominal detection threshold, and SNR=30. Orange stars represent known binaries detected through radio emission. grey horizontal bands represent chip masses of BNS detected to date through GW emission, GW170817 and GW190425.} 
\label{fig:snr_plot}
\end{figure*}

\begin{figure}
\centering
\includegraphics[width=1\columnwidth]{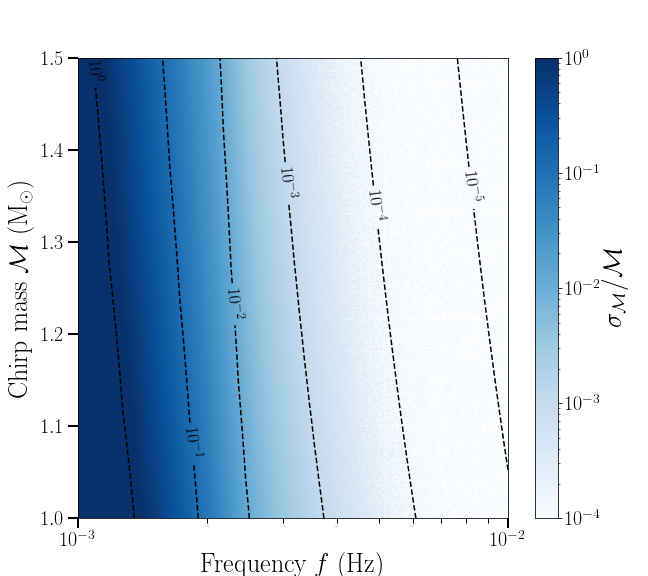}
\caption{Expected fractional error on the chirp mass as a function of GW frequency (x-axis) and chirp mass (y-axis) for a circular binary. Here we fixed the distance to the binary to $d=10\,$kpc and the observation time with LISA to $T=4\,$yr. Overlaid are lines of constant fractional measurement error on the chirp mass; their slope indicates that at a given GW frequency, higher chirp masses lead to a smaller error on the measurement.} 
\label{fig:errorMchirp}
\end{figure}

To start, we have to quantify the accuracy within which LISA can measure the chirp mass.
The chirp mass of a binary system is defined as
\begin{equation}
{\cal M}=\frac{(m_1 m_2)^{3/5}}{(m_1+m_2)^{1/5}},
\end{equation}
where $m_1$ and $m_2$ are the primary and secondary neutron star masses. 
It determines how fast the binary's GW frequency $f=2/P$ (with $P$ being binary orbital period) changes during the in-spiral phase
\begin{equation}
\dot{f}=\frac{96}{5}  \pi^{8/3} \left( \frac{G{\cal M}}{c^3} \right)^{5/3} f^{11/3},
\end{equation}
where $G$ and $c$ are respectively the gravitational constant and the speed of light.
Therefore, chirp mass can be derived from GW data when both frequency ($f$) and its time derivative ($\dot{f}$) are measured. The limiting frequency allowing the chirp mass measurement for a typical BNS is of $\sim 1.75\,$mHz (cf. Fig.~\ref{fig:errorMchirp}). At lower frequencies, BNS will be seen as monochromatic over the whole duration of the mission, meaning that their chirp mass will be degenerate with the distance (cf. Eq.~\ref{eqn:amp}).

In the case of a circular binary, the measurement of the chirp depends only on $f$ and $\dot{f}$.
The fractional error on the chirp mass can be estimated as 
\begin{equation} \label{eqn:sigmaMe0}
\frac{\sigma_{\cal M}}{{\cal M}} \simeq \frac{11}{5} \frac{\sigma_f}{f} + \frac{3}{5} \frac{\sigma_{\dot{f}}}{\dot{f}},
\end{equation}
where
\begin{equation} \label{eqn:sigmaF}
\frac{\sigma_f}{f} = 8.7 \times 10^{-7} \left( \frac{f}{2\,{\rm mHz}} \right)^{-1} \bigg( \frac{\rho}{10} \bigg)^{-1} \left( \frac{T}{4\,{\rm yr}} \right)^{-1}
\end{equation}
and 
\begin{equation} \label{eqn:sigma_fdot}
\frac{\sigma_{\dot{f}}}{\dot{f}} = 0.26 \left( \frac{f}{2\,{\rm mHz}} \right)^{-11/3} \left(  \frac{{\cal M}}{1.2\,{\rm M}_\odot} \right)^{-5/3} \bigg( \frac{\rho}{10} \bigg)^{-1} \left( \frac{T}{4\,{\rm yr}} \right)^{-1} 
\end{equation}
with $\rho$ being the signal-to-noise ratio for the mission time $T=4\,$yr \citep[e.g.][]{lau20}.
Averaging over sky location, polarisation, and inclination, one can write down the signal-to-noise ratio as \cite[e.g.][]{Robson_2019}:
\begin{equation} \label{eqn:snr}
    \rho^2 = \frac{24}{25} |{\cal A}|^2 \frac{T}{S_{\rm{n}}(f) R(f)},
\end{equation} 
where ${\cal A}$ is the amplitude of the signal 
\begin{equation} \label{eqn:amp}
    {\cal A} = \frac{2(G{\cal M})^{5/3}(\pi f)^{2/3}}{c^4 d},
\end{equation}
$S_{\rm{n}}(f)$ is the power spectral density of the detector noise in the low-frequency limit that also accounts for unresolved Galactic background, $d$ is the luminosity distance to the binary and $R(f)$ is a transfer function encoding finite-arm length effects at high frequencies that computed numerically in \citet{kor20}.  

Figure~\ref{fig:snr_plot} shows the signal-to-noise ratio (cf. Eq.~\ref{eqn:snr}) for a circular BNS placed at the distance of $10\,$kpc and observed over the nominal four years of the LISA mission. For comparison, we show the sub-sample of shortest orbital period Galactic BNSs detected through the radio emission (orange stars) with measured masses \citep{fer14,vanL15,kra16,cam18,sto18,far19}. By plugging in their true distances in Eq.~\eqref{eqn:snr}, we find that all of the known BNS lay under the LISA detection threshold of $\rho =7$. 
The chirp masses of extra-galactic BNS mergers detected through GW emission by LVC are delimited by grey horizontal bands. 
In Fig.~\ref{fig:errorMchirp} we zoom-in on frequencies > 1\,mHz and show the expected fractional error on the chirp mass. 

The eccentricity is another binary parameter that -- when sufficiently high -- can impact BNS detectability and, consequently, the chirp mass measurement.
In isolated binary evolution, BNSs are expected to form with non-zero eccentricity due to the supernova kicks associated with the formation of the last-born neutron star or Blaauw kicks produced by symmetric mass loss accompanying supernovae \citep{bla61,tau17}.
The GW radiation quickly circularises BNS orbits so that they become almost circular  by the time they move to the LISA band (e.g. when the Hulse-Taylor pulsar will evolve to 2\,mHz its eccentricity will decrease from 0.61 to 0.03). However, if the binary is born right in or at the edge of the LISA's frequency window, it will retain the original eccentricity. 
To assess how the eccentricity influences the chirp mass measurement and, consequently, LISA's ability to constrain the mixing fraction of a bi-modal chirp mass distribution we will explore two limiting cases: the case in which all binaries are circular and the case in which all binaries are eccentric.  
We anticipate that assuming all binaries to be eccentric does not significantly change our results on the mixing fraction. Thus, we defer the description of how the chirp mass measurement changes in the eccentric case to Appendix \ref{sec:A}.


\subsection{Mock Galactic BNS population} \label{sec:mockBNS}

\begin{figure}
\includegraphics[width=\columnwidth]{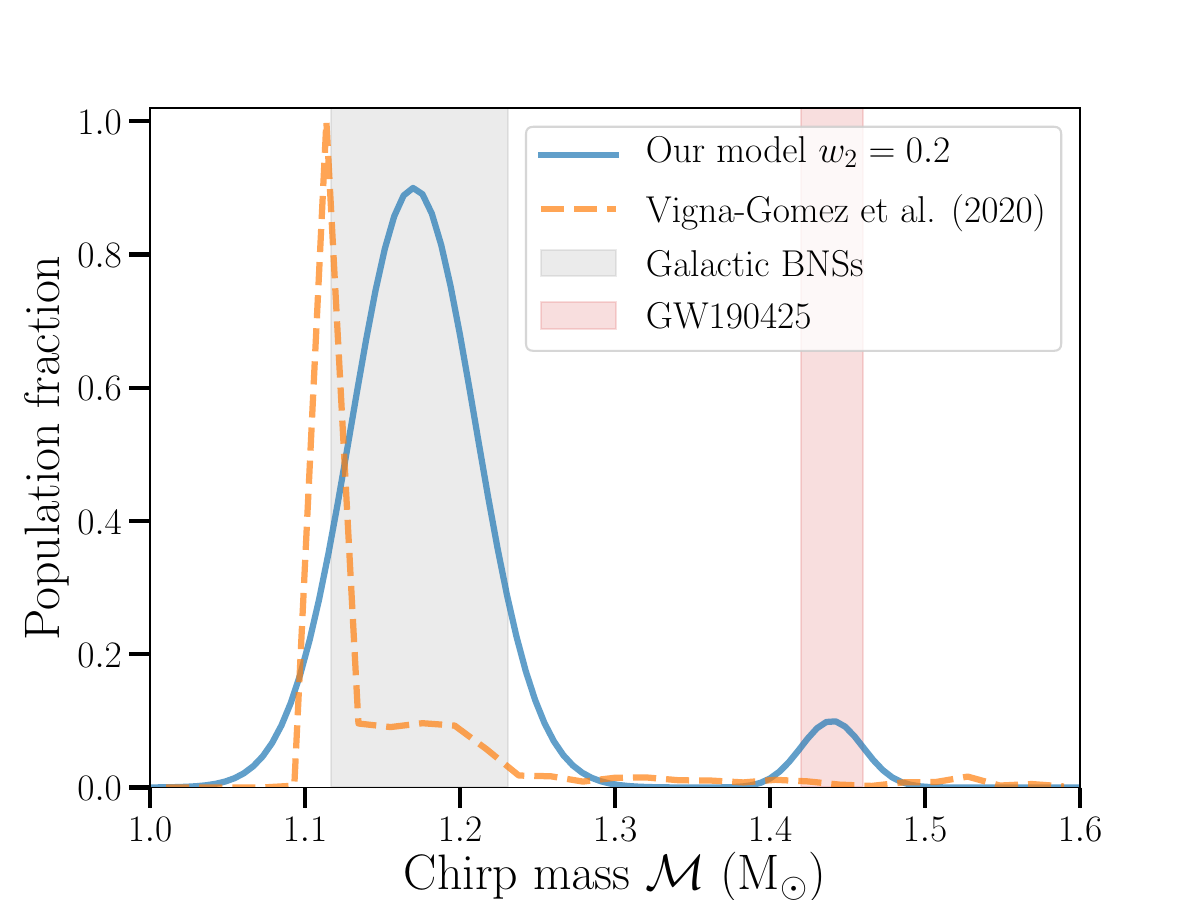}
\caption{Chirp mass distribution of our mock BNS population (blue solid). For comparison we show the chirp mass distribution from \citet{vig20} formed at $Z=0.0142$ (orange dashed). Vertical  shaded bands represent the range of chirp masses of known Galactic BNS in grey and that of GW190425 in red colours.} 
\label{fig:Mchirp_distribution}
\end{figure}

To assemble a mock Galactic population we assume that the BNSs' chirp mass distribution is described by a mixture of two Gaussian distributions: one centered on the value characteristic of the known Galactic population that has been detected at radio wavelengths and another centered on the chirp mass of GW190425. We therefore write
\begin{align} \label{eqn:p_pop}
P_{\rm pop}({\cal M})= &\frac{w_1}{\sigma_1 \sqrt{2\pi}} \exp \left(\frac{-({\cal M}-\mu_1)^2}{2\sigma_1^2}  \right)+ \nonumber\\
&\frac{w_2}{\sigma_2 \sqrt{2\pi}} \exp \left(\frac{-({\cal M}-\mu_2)^2}{2\sigma_2^2}  \right),
\end{align}
where $w_1$ and $w_2$ are the relative weights of the two Gaussian distributions normalised such that $w_1 + w_2 = 1$.
We obtain $\mu_1 = 1.17\,$M$_\odot$ with a standard deviation of $\sigma_1 = 0.04$ by combining individual chirp mass measurements of known Galactic BNS reported in \citet{far19}.  
We set $\mu_2 = 1.44\,$M$_\odot$ and $\sigma_2 = 0.02$ according to the posterior distribution reported in \citet{gw190425}. We show our model chirp mass distribution in Fig.~\ref{fig:Mchirp_distribution} with the blue solid line. 
We note that our choice is supported by studies analysing available GW and/or radio observations of BNSs that found evidence for a broad secondary peak at high masses in the birth mass distributions of second-born neutron stars \citep{far19,gal20}. In addition, population studies of binary white dwarf detectable with LISA also show bi-modality in the chirp mass distribution \citep[e.g.][]{kor17}.

To model the frequency distribution we assume that the Galactic BNSs population is stationary on
the time-scale of interest. Consequently, their distribution in frequency is
given by
\begin{equation} \label{eqn:dNdf}
\frac{dN}{df} = \frac{5c^5 {\cal R}_{\rm MW} }{96 \pi^{8/3} (G{\cal M})^{5/3} f^{11/3}}
\end{equation}
where ${\cal R}_{\rm MW}$ is the merger rate of BNSs in the Milky Way.
Note, however, that this assumption does not account for possible significant recent star formation episodes that could add BNS systems directly in the LISA band. 
We can then compute the total number of BNS at frequencies above $f$ by integrating Eq.~\eqref{eqn:dNdf}
\begin{align} \label{eqn:nbns}
 N(>f) &= \frac{5 c^5  {\cal R}_{\rm MW}}{256 \pi^{8/3} ( G\mathcal{M}_c )^{5/3} f^{8/3}} & \nonumber \\
 & \simeq33 \left( \frac{\mathcal{M}_c\,}{1.2 M_\odot} \right)^{-5/3} \left(
 \frac{f}{{2\,}{\rm mHz}} \right)^{-8/3} 
 \left(
 \frac{{\cal R}_{\rm MW}}{{140\,}{{\rm Myr}^{-1}}} \right),
\end{align}
where we use  ${\cal R}_{\rm MW} = 140$\,Myr$^{-1}$ \citep{set19}. 
We note that the Galactic BNS merger rate is still uncertain:
based on extra-galactic BNS merger rates reported by LVC in the first two observing runs \citet{and20} arrives at $210$\,Myr$^{-1}$, \citet{pol19} arrives $42$\,Myr$^{-1}$ using available radio observations, and \citet{vig18} predicts at $24$\,Myr$^{-1}$ based on the 
\texttt{COMPAS} binary population synthesis code. 
We note that the updated merger rate of $320^{+490}_{-240}$\,Gpc$^{-3}$yr$^{-1}$ based on the second LVC GW Transient Catalog \citep{GWTC2} corresponds to $32^{+49}_{-24}$\,Myr$^{-1}$ (assuming the number density of the Milky Way-like galaxies of $0.01$\,Mpc$^{-3}$), and thus agrees better with the rates estimated from the population of Galactic BNSs.
Moreover, in the next decades, the differences in estimated merger rates should further decrease as more observations will become available from both radio and GW observatories.

Finally, we assume that BNS are distributed in the Galactic disc with an exponential radial stellar profile with an isothermal vertical distribution
\begin{equation} \label{eqn:bns_positions}
P(R,z) \propto e^{-R/R_{\rm d}} \sech^2(z/z_{\rm d})
\end{equation}
where $0 \le R \le 20$\,kpc is the cylindrical radius measured from the Galactic centre, $z$ is the height above the Galactic plane, $R_{\rm d} = 2.5\,$kpc is the characteristic scale radius, and  $z_{\rm d}=0.4$\,kpc is the vertical scale height of the observed Galactic BNSs \citep{pol19}. 
Finally, when converting BNS positions $(R,z)$ into heliocentric distances $d$ we assume the position of the Sun to be at $(8.1,0)$ according to \citet{abu19}.



\subsection{Bayesian inference}

Given $N$ measurements of the BNS chirp masses with associated errors, we now would like to reconstruct the shape of the underlying true chirp mass distribution.
Here we have chosen to model the chirp mass of Galactic BNSs as a mixture of two Gaussian distributions (cf. Eq.~\ref{eqn:p_pop}), thus the distribution is fully described by 6 parameters 
$\lambda \in \{w_1, w_2, \mu_1, \mu_2, \sigma_1, \sigma_2\}$.
We follow the Bayesian approach as outlined in \citet{man19} ignoring selection effects because we are only interested in a sub-sample of detections with $f>2\,$mHz -- that allows the measurement of the chirp mass --  across which we find the detection probability with LISA to be $\sim 1$. 

To simulate the instrumental noise, we displace each chirp mass from its true value by re-sampling it from a Gaussian centered on the true value and standard deviation determined by the LISA's measurement error $\sigma_{\cal M}$ (cf. Eq.~\ref{eqn:sigmaMe0}). We will denote this displaced chirp mass with $\hat{{\cal M}}$.

Using Bayes' theorem, we can write the posterior as
\begin{equation}  \label{eqn:posterior}
P(\lambda|\hat{{\cal M}}) \propto \pi(\lambda) \prod_{i=1}^{N} \int d{\cal M}_i P(\hat{{\cal M}_i}|{\cal M}) P_{\rm pop}({\cal M}_i|\lambda),
\end{equation}
where $\pi (\lambda)$ are priors on $\{w_1, w_2, \mu_1, \mu_2, \sigma_1, \sigma_2\}$, $P(\hat{{\cal M}_i}|{\cal M})$ is the probability of observing the event $i$ given the assumed underlying distribution (likelihood), $P_{\rm pop}$ is the probability of individual chirp mass given the underlying population distribution (Eq.~\ref{eqn:p_pop}). We assume that the likelihood follows a Gaussian distribution. We adopt uniform priors for $w_1, w_2 \in [0,1]$, $m_1 \in [1 - 1.3]\,$M$_{\odot}$ and for $m_2 \in [1.3 - 1.5]\,$M$_{\odot}$, and a log-uniform prior for $\sigma_1, \sigma_2 \in [-3,-1]$.

\section{Results}

\begin{figure*}
\includegraphics[width=1.85\columnwidth]{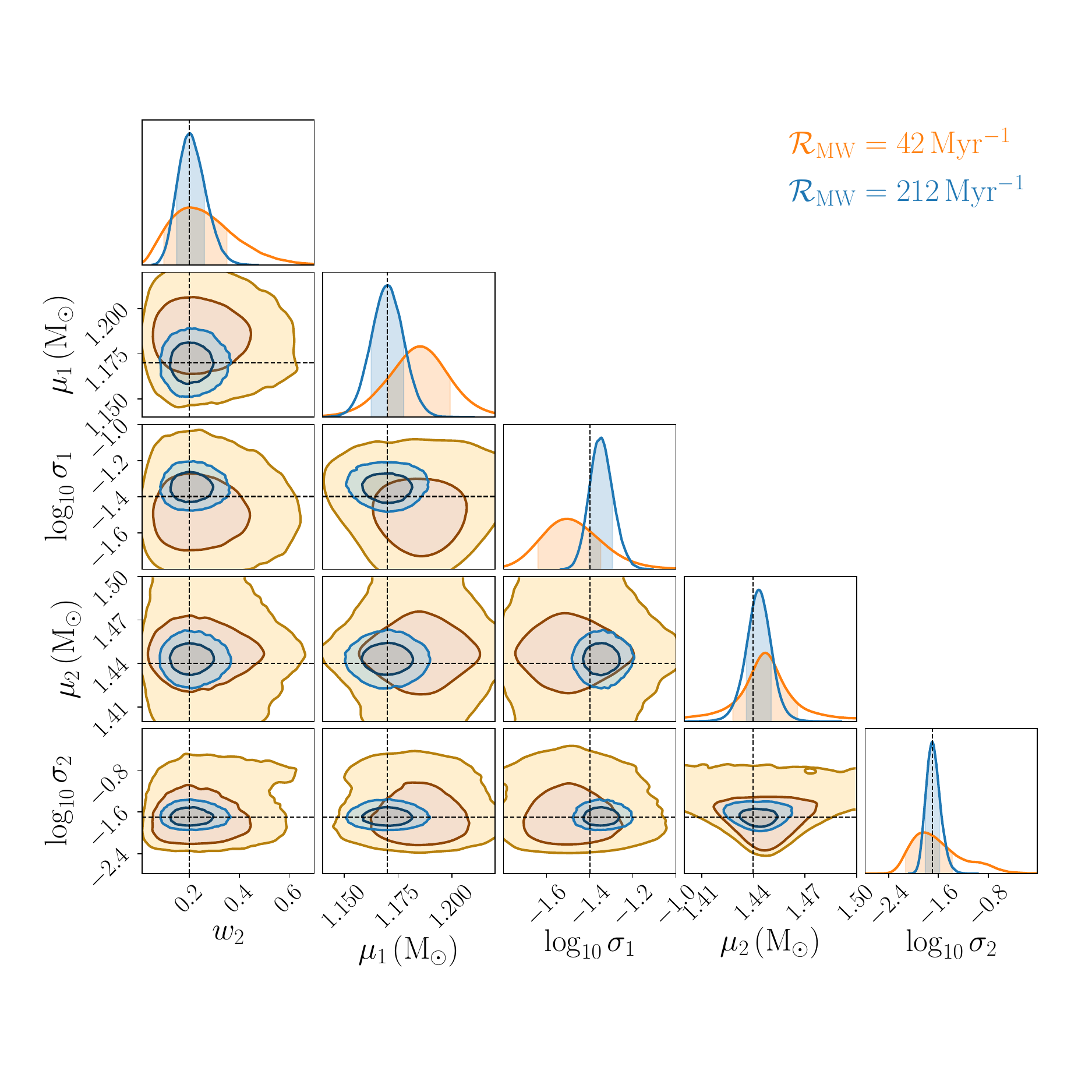}
\caption{Results of the MCMC simulation in recovering input parameters of the adopted model with $w_2=0.2$ and the other four variables that enters Eq.~\eqref{eqn:p_pop}: for ${\cal R}_{\rm MW} = 42\,$Myr$^{-1}$ in yellow and for ${\cal R}_{\rm MW} = 212\,$Myr$^{-1}$ in blue. The true values are marked by black dashed lines. The dark and light shaded regions indicate the 68 per cent, and 95 per cent confidence region in the derived posteriors. The top left panel shows that the fraction of GW190425-like binaries is constrained in both cases. } 
\label{fig:posterior}
\end{figure*}

We start by assuming the fraction of BNSs that resemble GW190425 to be $w_2=0.2$ and we set the Galactic merger rate to be $140\,$Myr$^{-1}$, which corresponds to 33 BNSs with frequencies higher than 2\,mHz in the Galaxy (cf. Eq.~\ref{eqn:nbns}). 
We recover the parameters of the chirp mass distribution in the framework of probabilistic programming package \texttt{pyMC3} \citep{pymc3} and sample the posterior (Eq.~\ref{eqn:posterior}) using the No U-turn Sampler. 
Figure \ref{fig:posterior} shows two examples of posterior distributions for $w_2, \mu_1, \mu_2, \sigma_1$ and $\sigma_2$.  
We have chosen to display only one of the two weights $w_2$, since the other can be recovered as $1-w_2$.
We present our results for two assumptions on the BNS merger rate: ${\cal R}_{\rm MW} = 42\,$Myr$^{-1}$ in orange and ${\cal R}_{\rm MW} = 212\,$Myr$^{-1}$ in blue colours. The true values are marked by black dashed lines. It is immediately evident that the higher merger rate leads to more constrained results because the number of detectable BNSs by LISA is higher. Moreover, in the case of the lower merger rate the small number statistics leads to bias in some of the parameters, although the true value is recovered to within the $1-\sigma$ of the derived posteriors. Importantly, Fig.~\ref{fig:posterior} shows that we can constrain the relative weights of the two BNS sub-populations $w_1$ and $w_2$.

We now focus on how well our inference machinery can recover $w_2$ (or equivalently $w_1$). We fix the Galactic merger rate and vary $w_2$. We find that for ${\cal R}_{\rm MW} = 140\,$Myr$^{-1}$ the weight of GW190425-like sub-population is recovered with $1\sigma$ error of 0.1 -- 0.04,
where the largest error is obtained for $w_2=0.1$ while the smallest for $w_2=0.9$.
This trend can be explained by remembering that the error on the chirp mass is mainly dominated by $\sigma_{\dot{f}} \propto {\cal M}^{-5/3}$ (cf. Eq.~\ref{eqn:sigma_fdot} for the circular case; for the eccentric case this is true only up to $\sim 3\,$mHz where $\sigma_{\dot{f}}$ and $\sigma_e$ start to be comparable, see Appendix \ref{sec:A}). Thus, the population with $w_2=0.9$ has overall better measured chirp masses than a population with $w_2=0.1$. 
Next, we fix $w_2$ and vary the merger rate in the range between 21 -- 212\,Myr$^{-1}$. We find that the error on the recovered fraction decreases with increasing merger rate as more LISA observations become available. Specifically, when setting the `true' value $w_2=0.2$ we recover $0.34^{+0.34}_{-0.22}$ for ${\cal R}_{\rm MW} = 21\,$Myr$^{-1}$, $0.22^{+0.08}_{-0.07}$ for ${\cal R}_{\rm MW} = 140\,$Myr$^{-1}$, and $0.19^{+0.06}_{-0.05}$ for ${\cal R}_{\rm MW} = 212\,$Myr$^{-1}$.

\begin{figure}
\includegraphics[width=\columnwidth]{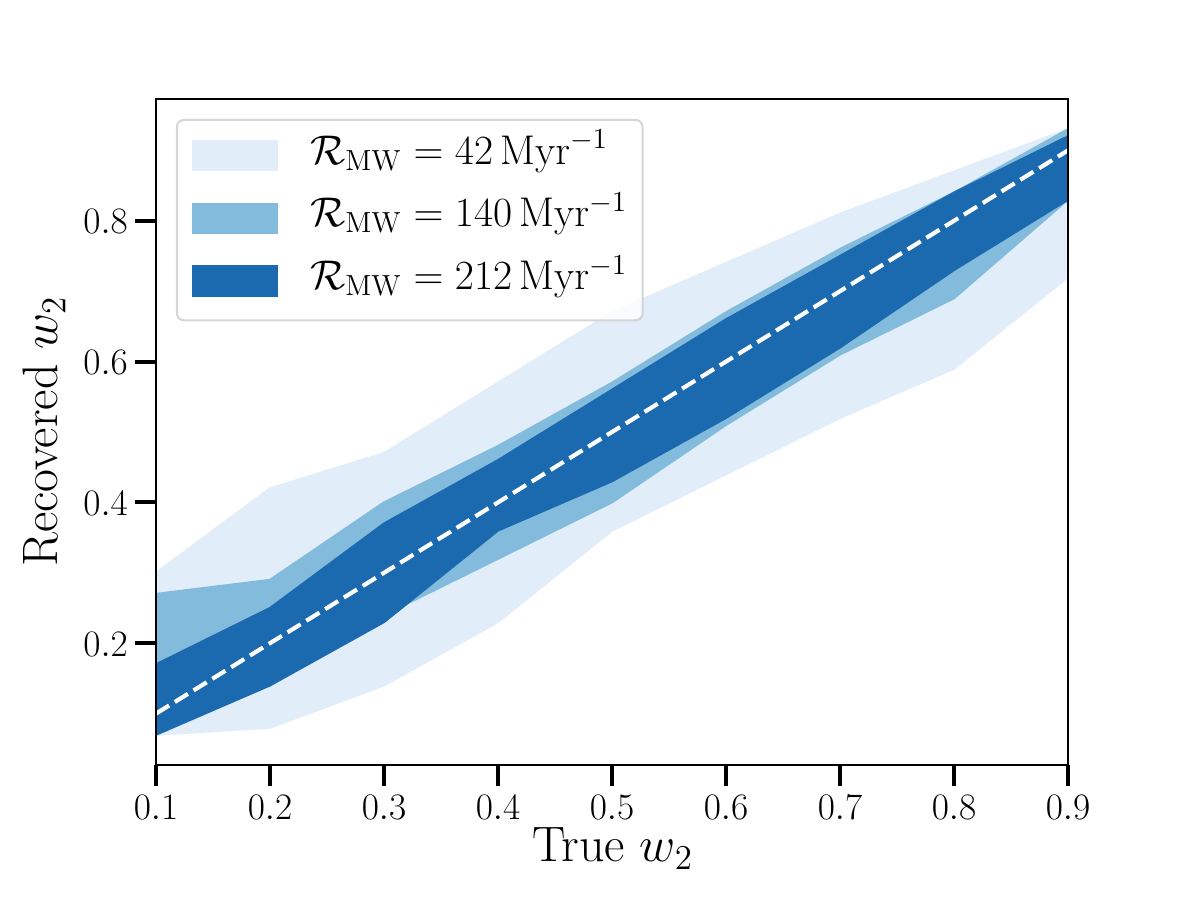}
\caption{Shows the recovered input fraction of GW190425-like BNSs in our mock population ($w_2$) as a function of the true fraction at fixed merger rates of 42, 140 and 212\,Myr$^{-1}$.
The blue band shows the $1-\sigma$ dispersion around the recovered fraction. The dashed line shows the 1-to-1 relation.} 
\label{fig:5}
\end{figure}

The summary of the recovered $w_2$ as a function of the `true' (input) $w_2$ is represented in Fig.~\ref{fig:5}. It shows that chirp mass measurement errors allow the recovery of the two sub-populations' weights with no bias regardless of its `true' value.
In colour we show our ability to recover $w_2$ for different Galactic merger rates of $42, 140, 212\,$Myr$^{-1}$, which corresponds to the total number of observed binaries at $f>2\,$mHz of $10, 33, 50$ respectively (cf. Eq.~\ref{eqn:nbns}). 

Finally, for comparison, we perform the same set of simulations as presented above for the limiting case in which all BNSs in the LISA band are eccentric. For example, using different natal kick prescriptions \citet{lau20} reports median eccentricity of the population ranging from 0.36 to 0.071, with a median of 0.1 for their fiducial model. For simplicity, as an example here, we set the eccentricity of all binaries to 0.3. We do not find any significant deviations in re-covering $w_2$ from the results of the circular case presented above. We can attribute this to the fact that the chirp mass error distribution does not differ significantly between the circular and eccentric cases (see also Appendix \ref{sec:A}).
 
\section{Discussion and conclusions}

\begin{figure}
\includegraphics[width=\columnwidth]{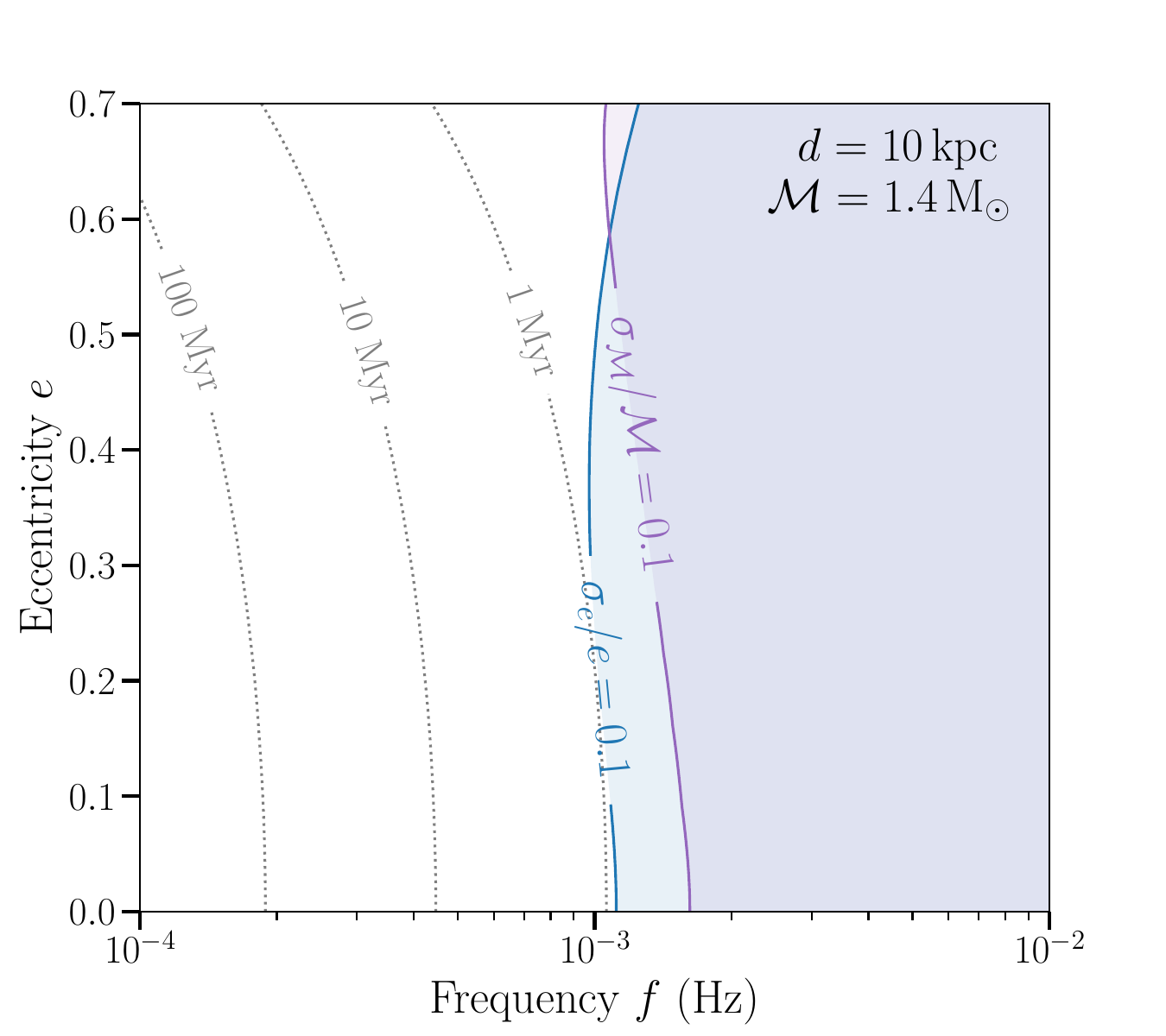}
\caption{Expected fractional error of 0.1 on the chirp mass (in purple) and the eccentricity (in blue) for a BNS with the chirp mass of ${\cal M}=1.4\,$M$_{\odot}$ at the distance of $d=10$\,kpc. The overlapping region shows where both quantities are measured and thus where constrains on the origin of GW190425-like BNSs are possible. Contours of the merger times of 1, 10 and 100\,Myr are marked by the grey dotted lines.} 
\label{fig:6}
\end{figure}

In this paper, we have investigated whether future observations of Galactic BNS with LISA and specifically LISA's chirp mass measurement can elucidate on the nature of heavy BNS like GW190425. First, we showed that  if GW190425-like binaries populate frequencies $> 2\,$mHz, they can be detected by LISA within the volume of our Galaxy. Moreover, their chirp masses can be measured to better than 10\,per cent. Then, we constructed a toy model for Galactic BNSs consisting of two distinct sub-populations: one that follows the chirp mass distribution of known BNSs detected through radio observations and another that resembles GW190425. 
Since the relative fraction of the two sub-population depends on the BNS formation model and the treatment of the fast-merging channel in modeling the total population in the Galaxy  \citep[e.g.][]{vig18,2021MNRAS.500.1380M}, here we considered the weights of the two sub-population as free parameters. As the Galactic BNS merger rate is still uncertain, we repeat our study for a range of merger rates between $24 - 212$\,Myr$^{-1}$ quoted in the literature \citep{gw170817,vig18,pol19,and20}.
We demonstrated that if GW190425-like binaries constitute a fraction larger than $0.1$ of the total population, LISA should be able to recover the fraction with better than $\sim15\,$per cent accuracy assuming the merger rate of ${\cal R}_{\rm MW} = 42\,$Myr$^{-1}$ (corresponding to 10 detected binaries with $f> 2\,$mHz); the accuracy increases to $\sim 5\,$per cent for  ${\cal R}_{\rm MW} = 212\,$Myr$^{-1}$ (corresponding to 50 detected binaries with $f> 2\,$mHz).
We note that with more upcoming radio and GW observations, the BNS merger rate is expected to be better constrained, such that we will have a more precise estimate on the expected number of BNS sources detectable by LISA. The results of this work can then be used to access what fractions of more massive GW190425-like binaries can be constrained by LISA.
Finally, our results show that even if all Galactic BNSs come with large eccentricities, then the errors on the measured chirp masses stay within the limits where our recovered fractions, assuming circularized orbits, remain applicable (cf. Fig.~\ref{fig:6}).

Further considerations on BNS formation could be deduced from the comparison of the BNSs rate determined from radio observations of pulsars at sub-mHz frequencies (stars in Fig.~\ref{fig:snr_plot}) with that inferred from LISA at mHz frequencies.
If all BNS form at low frequencies (merger times of $> 100\,$Myr) and evolve to mHz frequencies via GW emission (cf. Eq.~\ref{eqn:nbns}), the two rates should be consistent. Whereas if the BNS rate at mHz frequencies is larger than predicted by Eq.~\eqref{eqn:nbns}, the excess could be attributed either to a sub-population of radio-quite BNSs \citep[e.g.][]{saf20} or to the fast-merging channel.

Population synthesis studies suggest a broad distribution of delay times between BNS formation and merger even under the assumption that case-BB mass transfer is dynamically stable \citep[e.g.][]{tau13,tau15,vig18}.
However, if the case-BB mass transfer is occasionally dynamically unstable \citep{dew03,iva03}, it is possible that fast-merging BNSs form within the LISA band at higher frequencies than the rest of the Galactic population. Therefore, if their merger timescales are sufficiently long to be observed with LISA, it is possible that fast-merging BNSs exhibit an excess in the frequency distribution.
LISA's frequency measurement is expected to be very accurate (ideally $\sigma_f/f \propto 1/T \sim 10^{-8}$ for $T=1\,$yr), thus identifying a pile-up of BNSs at high frequencies should be a straight forward task. Importantly, the observed frequency distribution will facilitate constraints on the BNS delay time distribution.
Moreover, high frequency binaries would also have better chirp mass measurements (and other parameters including the eccentricity) that could only improve the results presented here.

To confidently constrain the fast-merging hypothesis eccentricity measurements are required \citep{and19,and20,lau20}.
If high mass GW190425-like binaries are also highly eccentric ($e \sim 1$), this would imply that they are freshly formed (either in isolation or in a dense cluster), and thus come from the fast-merging channel. Figure~\ref{fig:6} shows that constraining both the chirp mass and the eccentricity to, for example, 10\,per cent is only possible at mHz frequencies.
\citet{and20} showed that such fractional errors on the eccentricity should be sufficient to distinguish between the different BNS formation channels. Still, if the eccentricity of GW190425-like binaries is low ($e \sim 0$) one cannot rule out the possibility that they are a population radio-quite BNSs.

We remark that our error estimates are based on the Fisher information matrix approximation and are valid for high signal to noise ratios \citep[e.g.][]{cut98}. Thus, we expect that in some cases the reported errors may be underestimated. A full Bayesian parameter estimation would be required to derive more realistic uncertainties \citep[e.g.][]{bus19,roe20}.

In this worked we focused exclusively on the LISA mission, but our analysis would hold also for the TianQin \citep{luo16,mei20} 
and Taiji \citep{rua18} space-based GW observatories designed to operate in a similar frequency range as LISA. \citet{hua20} showed that LISA and TianQin can simultaneously detect Galactic stellar binaries above a few mHz and that combined observations from the two missions will improve their parameters estimation including orbital period, inclination and sky localisation.

While radio observatories are sensitive to the intermediate stages of BNS evolution and ground-based GW detectors can see only the very last seconds of a BNS's life and the final merger, LISA will be crucial for bridging the gap between these two regimes \citep{gal20}.
Here we argued that the BNS chirp mass distribution measured by LISA could be a useful tool for unveiling massive GW190425-like BNSs in the Milky Way and constraining their origin.
Other studies established the importance of the LISA's eccentricity measurement for understanding the BNS formation pathways \citep{and20,lau20}. 
In addition, LISA will also provide sky positions for Galactic BNS enabling radio follow-up and the discovery of radio-faint pulsars with all-sky surveys such as the Square Kilometre Array \citep{kyu19}.

\section*{Acknowledgements}

We thank Silvia Toonen, Alejandro Vigna-G{\'o}mez, Ilya Mandel, Riccardo Buscicchio, Antoine Klein, Davide Gerosa, Christopher Moore and Alberto Vecchio for useful discussions and suggestions.
VK acknowledges support from the Netherlands Research Council NWO Rubicon 019.183EN.015 grant. MS thanks the Heising-Simons Foundation, the Danish National Research Foundation (DNRF132), and NSF (AST-1911206 and AST-1852393) for support. This work made use probabilistic programming package \texttt{PyMC3} \citep{pymc3} and \texttt{ChainConsumer python} plotting package \citep{chainconsumer}.

\section*{Data Availability}

No new data were generated or analysed in support of this research.



\bibliographystyle{mnras}
\bibliography{vale} 



\appendix

\section{Eccentric case} \label{sec:A}

Differently from circular binaries, eccentric binaries emit GWs at multiple harmonics. Each harmonic can be thought as a collection of $n$ circular binaries emitting at $ f_n =n f/2$ and the amplitude ${\cal A}_n={\cal A}(2/n)^{5/3} g(n,e)^{1/2}$, where $g(n,e)$ is given in \citet{pet63}. Note that $g(2,0) = 1$ so we recover the circular case.
The total signal-to-noise ratio can be estimated as the quadrature sum of the individual harmonics' signal-to-noise ratios $\rho_n$. Thus, under the assumption of the optimal GW signal recovery, the signal-to-noise of an eccentric binary is always greater than that of a circular one.

Relevant to the chirp mass measurement, the frequency evolution in the eccentric case depends on both the chirp mass and the eccentricity
\begin{equation}
n\dot{f}=\frac{96}{5} \left(\frac{ 2\pi}{n} \right) ^{8/3} \left( \frac{G{\cal M}}{c^3} \right)^{5/3} (n f) ^{11/3} \, F(e),
\end{equation}
where $F(e)$ is the enhancement factor \citep{pet63}
\begin{equation} \label{eqn:fe}
F(e) = \frac{1+\frac{73}{24}e^2+\frac{37}{96}e^4}{(1-e^2)^{7/2}}.
\end{equation}
Thus, Eq.~\ref{eqn:sigmaMe0} now becomes
\begin{equation} \label{eqn:sigmaMe}
\frac{\sigma_{\cal M}}{{\cal M}} \simeq \frac{11}{5} \frac{\sigma_f}{f} + \frac{3}{5} \frac{\sigma_{\dot{f}}}{\dot{f}} + \frac{3}{5}\frac{\sigma_{F(e)}}{F(e)}.
\end{equation}
Consequently, in the eccentric case to measure the chirp mass one needs to simultaneously measure the eccentricity, otherwise only upper limit of the chirp mass can be derived.
The relative ratio of any two harmonic is proportional to the eccentricity. Therefore, detection of at least two harmonics is required to measure binary's eccentricity \citep[e.g.][]{seto16}.
For relatively moderate eccentricities that we have considered in this work ($e\leq 0.3$), $f_2$ and $f_3$ are the strongest harmonics and
\begin{equation}
\frac{\sigma_e}{e} \simeq \left( \frac{1}{\rho_2^2} + \frac{1}{\rho_3^2} \right)^{-1/2}.
\end{equation} 
Finally, the contribution to chirp mass error due to eccentricity $\sigma_{F(e)}$ can be
calculated directly for known $\sigma_e$ using Eq.~\eqref{eqn:fe}.

\begin{figure}
\centering
\includegraphics[width=0.8\columnwidth]{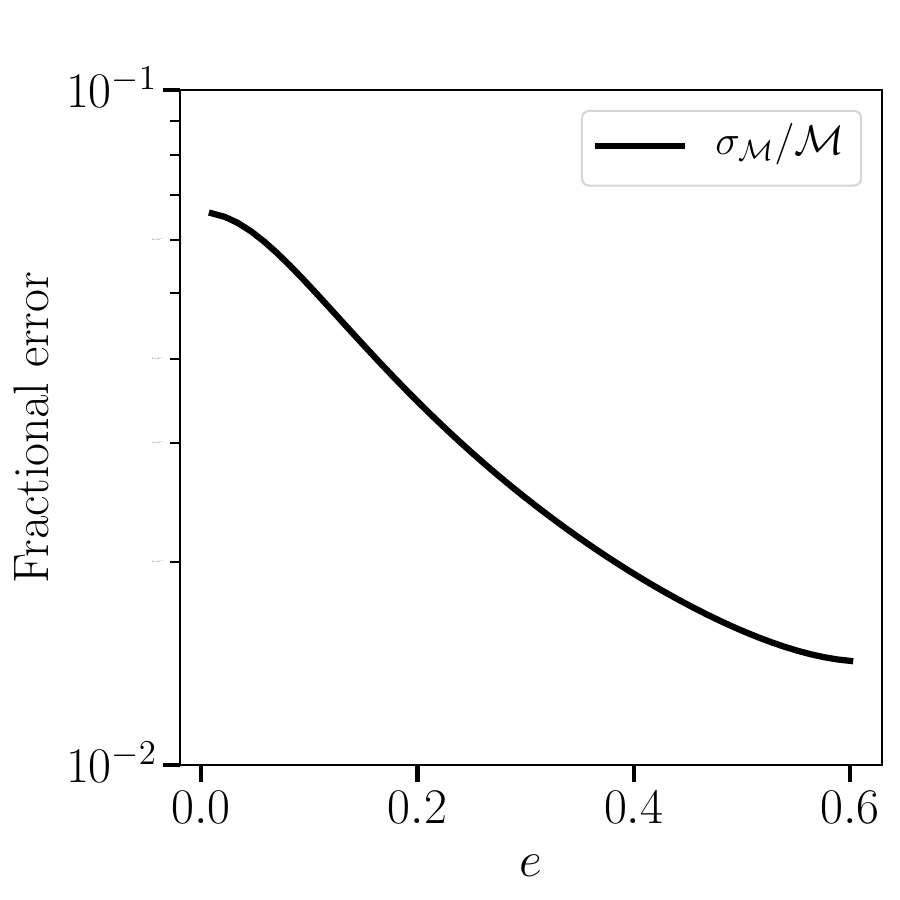}
\includegraphics[width=0.8\columnwidth]{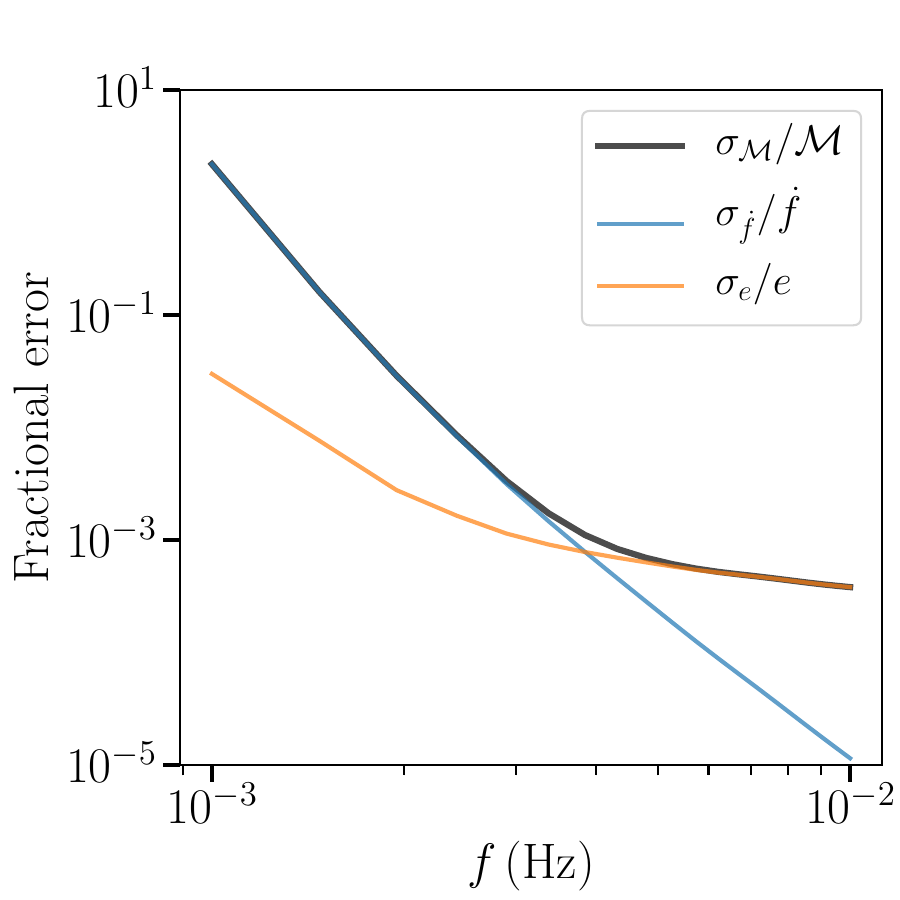}
\caption{{\it Top panel}: Fractional error on the chirp mass as a function of the eccentricity for a binary at the distance of $d=10\,$kpc and emitting at 2\,mHz. {\it Bottom panel:} Fractional error of the chirp mass (black) for a binary with $e=0.3$ split into the contributions due to $\dot{f}$ and $e$; we do not represent the contribution due to the uncertainty in $f$ as it remains negligible at all frequencies. } 
\label{fig:errorMchirp_e}
\end{figure}

In the top panel of Fig.~\ref{fig:errorMchirp_e} we show how the fractional error on the chirp mass changes as a function of the eccentricity for a binary at the distance of $d=10\,$kpc and emitting at 2\,mHz.
In the bottom panel of Fig.~\ref{fig:errorMchirp_e} we fix binary's eccentricity to $e=0.3$ and split fractional error of the chirp mass (black) into the contributions due to $\dot{f}$ (blue) and $e$ (orange); we do not represent the contribution due to the uncertainty as it remains negligible at all frequencies. For $f \lesssim 3\,$mHz the error on the chirp mass is dominated by the error on $\dot{f}$, which scales with $\rho$ and thus $\sigma_{\cal M}/{\cal M}$ is smaller compared to in the circular case (see upper panel).
At $3-4\,$mHz, the contribution of the eccentricity limits the chirp mass measurement, so at higher frequencies the chirp mass is better for the circular binaries.

\bsp	
\label{lastpage}
\end{document}